# Selective particle trapping and optical binding in the evanescent field of an optical nanofiber

**M. C. Frawley,[1,2] I. Gusachenko,[1] V. G. Truong,[1*] M. Sergides,[1] and S. Nic Chormaic[1]**

[1]Light-Matter Interactions Unit, OIST Graduate University, 1919-1 Tancha,
Onna-son, Okinawa 904-0495, Japan
[2]Physics Department, University College Cork, Cork, Ireland
* v.g.truong@oist.jp

**Abstract:** The evanescent field of an optical nanofiber presents a versatile interface for the manipulation of micron-scale particles in dispersion. Here, we present a detailed study of the optical binding interactions of a pair of 3.13 μm $SiO_2$ spheres in the nanofiber evanescent field. Preferred equilibrium positions for the spheres as a function of nanofiber diameter and sphere size are discussed. We demonstrated optical propulsion and self-arrangement of chains of one to seven 3.13 μm $SiO_2$ particles; this effect is associated with optical binding via simulated trends of multiple scattering effects. Incorporating an optical nanofiber into an optical tweezers setup facilitated the individual and collective introduction of selected particles to the nanofiber evanescent field for experiments. Computational simulations provide insight into the dynamics behind the observed behavior.

## 1. Introduction

In 1986, Ashkin et al. demonstrated the first 'optical tweezers,' whereby a tightly focused laser beam was used to trap colloidal particles. Such a beam creates an optical force, stably trapping dielectric particles at the beam waist, and enabling three dimensional positional motion via movement of the beam [1]. Later, an intense laser beam was used to create a long, narrow focus in which colloidal particles were trapped in ordered arrays [2]. These were formed due to the scattering effects of the laser light. This phenomenon is termed 'optical binding', and involves the modification of the incident light field in the presence of multiple, simultaneously illuminated objects, which mutually optically interact and self-arrange into ordered configurations. Here, a given particle induces light redistribution, resulting in attractive or repulsive forces, which in turn influence the position of neighboring particles.

The first evidence of longitudinal optical binding was presented by Tatarkova et al. in 2002 [3]. Using two weakly focused counter-propagating Gaussian beams, silica microspheres with diameters 2.3 μm and 3 μm were trapped along the propagation direction of the incident field. The beams were tailored so that the radiation pressure of each one was cancelled by the other. The particles were confined in equilibrium positions with spacings of several times the particle diameters.

The aforementioned work used free-space beams; the use of evanescent fields to optically manipulate particles near a surface provides an alternative geometry with some advantages. The propulsion of microparticles in this manner was first demonstrated by Kawata and Sugiura in 1992 [4]. Microparticles were driven along the surface of a high refractive index prism by the radiation pressure of an evanescent wave, created by a single infrared laser beam. In this regime, the evanescent wave converts to a traveling wave within the particle and a fraction of the momentum component parallel to the surface is transferred to it.

Evanescent field effects can be observed in other optical structures, e.g. in standard optical fibers, although they are not normally significant. In cases where the size of the structure is comparable to the optical wavelength, such as in an optical 'nanofiber', the evanescent waves play a dominant role and an enhanced optical interaction with nearby structures can be observed [5]. In the past thirty years, optical nanofibers have demonstrated an abundance of useful applications as optical interface tools [6]. Typically produced in the laboratory by simply taper-drawing commercial optical fiber over a localized heat source, the submicron cross-section combines high mode confinement with a pronounced evanescent field which protrudes into the surrounding medium. This facilitates high sensitivity environmental and particle sensing [7] as well as atomic and molecular detection and spectroscopy [8]. The gradient profile of the evanescent field has also been exploited for experiments in micron-scale particle manipulation [9]. Recently, optical nanofibers have been used for propulsion, size-sorting and mass migration of glass, plastic, metallic, and biological particles in solution [10-12]. Such demonstrations have shown the potential of nanofibers for long-range optical trapping at arbitrary points in a given sample. Microfluidic-nanofiber experiments have also laid the foundations for broader applications in lab-on-a-chip systems [13].

Although optical binding close to planar surfaces has received substantial attention in recent years [14] the same cannot be said for binding interactions in the nanofiber evanescent field. Aside from analysis of single-particle trapping forces and speeds, the subtleties of the optical interaction of multiple particle systems and the nanofiber have been, thus far, largely overlooked. However, optical binding is often observed during particle propulsion and a better understanding of this is necessary for future practical applications of nanofibers as optical manipulation and trapping tools. The phenomenon of optical binding of particles in the nanofiber evanescent field is studied both theoretically and experimentally in this paper. The role of the Maxwell stress tensor in calculating the force on a particle in an optical field is highlighted in Section 2. Section 3 introduces simulations of the forces on particle pairs in the nanofiber vicinity over a range of inter-particle separations; the optical forces are decomposed into gradient and scattering components, and the long and short range optical binding interactions highlighted. Stable inter-particle distances are deduced by locating the positions of minimum potential. Additionally, the effects on equilibrium binding positions of varying sphere size and nanofiber diameter, and of transmission loss due to scattered light from the first particle, are considered.

To systematically investigate these interactions, a controlled method of selective particle introduction to the fiber is necessary. To achieve this, we combined custom-made nanofibers within the framework of an optical tweezers [15]. This experimental setup is briefly described in Section 4.1, and the optical propulsion of chains of one to seven 3.13 μm $SiO_2$ (silica) particles is demonstrated, indicating optical binding characteristics. Finally, typical experimental speeds of particle chains in the nanofiber evanescent field are presented and compared with the speed of a single particle in the same incident field.

## 2. Theory of force calculation

The distribution of electromagnetic waves can be determined from Maxwell's wave equation. Using a commercially-available finite element software package (COMSOL Multiphysics 4.3b), the spatial variation of the electromagnetic fields along the nanofiber and particles can be ascertained. An eigenvalue equation for the electric field $\boldsymbol{E}$ is derived from

$$\nabla \times (\nabla \times \boldsymbol{E}) - k_0^2 n^2 \boldsymbol{E} = 0, \tag{1}$$

where $k_0$ is the wavenumber and $n$ the material refractive index.

The total optical force acting on an object is calculated by integrating the Maxwell stress tensor on the surface of the particle. The stress-tensor $\langle \boldsymbol{T_M} \rangle$ is given by [16, 17]:

$$\langle \boldsymbol{T_M} \rangle = \boldsymbol{D}\boldsymbol{E}^* + \boldsymbol{H}\boldsymbol{B}^* - \frac{1}{2}(\boldsymbol{D}.\boldsymbol{E}^* + \boldsymbol{H}.\boldsymbol{B}^*)\boldsymbol{I}, \tag{2}$$

where $E$, $D$, $H$ and $B$ denote the electric field, the electric displacement, the magnetic field and magnetic flux, respectively. $I$ is the isotropic tensor, and $E^*$ and $H^*$ are the complex conjugates. Applying the constitutive relations $D = \varepsilon_r \varepsilon_0 E$ and $B = \mu_r \mu_0 H$, Eq. 2 can be rewritten as

$$T_{ij} = \varepsilon_r \varepsilon_0 E_i E_j^* + \mu_r \mu_0 H_i H_j^* - \frac{1}{2}(\varepsilon_r \varepsilon_0 E_k E_k^* + \mu_r \mu_0 H_k H_k^*)\delta_{ij}, \qquad (3)$$

where $E_{ij}$ and $H_{ij}$ are the $i^{th}$ components of the electric and magnetic fields; $\varepsilon_r$, $\varepsilon_0$, $\mu_r$ and $\mu_0$ are the relative and vacuum permittivity and permeability, respectively, $\delta_{ij}$ is the Kronecker delta function, and we sum over the indices $k$.

The optical force exerted on an arbitrary object is defined as

$$F = \oint_s (\langle T_M \rangle \cdot n_s)\, dS, \qquad (4)$$

where $n_s$ is the normal vector pointing outward from the surface $S$ of the object. For a spherical particle in the evanescent field of a horizontal optical nanofiber, the horizontal (scattering) and vertical (gradient) optical forces are calculated by integrating the external fields at the particle surface in this system.

Here, the total force includes the optical forces acting on the particle and on the water interface. This is transmitted to the sphere due to the hydrodynamic non-slip boundary conditions at the interface. Particles are assumed to position themselves where the intensity of the incident evanescent wave is strongest, hence the trapping gradient force acting on a particle is highest when particles are close to the surface of a nanofiber. For the following simulations, the distance between particles and the nanofibers is assumed to be 20 nm. For 1064 nm propagating light, $\lambda$, the refractive index of the fiber, water and silica particles are $n_{fiber} = 1.45591$, $n_{water} = 1.33$ and $n_{silica} = 1.46$, respectively. For all calculations, the propagating power through the nanofiber is normalized to 10 mW. To validate the numerical calculation of theoretical models, we compared the calculated optical forces using the surface integrated stress tensor approach to the integrated force density. For the latter, the integral of the force density is taken over the volume of the particle [16]. Assuming that the permittivity, $\varepsilon_r$, is constant in the volume of a sphere, the integral is therefore non-zero on the surface of the particle. The total integration of the Maxwell stress tensor can be then simplified to the surface integral. Analytically, both force calculation methods should give identical values. However, this is only valid for sufficiently high-resolution regimes. In our case the mesh size used in calculations is smaller than $\lambda/7$, where $\lambda$ is the wavelength of the propagating light.

### 3. Simulation

*3.1 Interactions between two microspheres in the nanofiber evanescent field*

Previous theoretical and experimental studies have investigated the one-dimensional longitudinal optical binding effect of microparticles in a counterpropagating geometry of Bessel light beams [21, 22]. The study of this binding effect was based upon the interference of the incident Bessel beam with far-field scattered spherical waves off particles. In this paper, we first provide numerical and quantitative experimental results on the longitudinal optical binding effect of self-assembled particle chains in the evanescent field of a nanofiber. The interference of the field in this case is due to the different wave vectors of the nanofiber-induced evanescent field, and the scattered field off particles along the optical axis of the nanofiber. The nanofiber guided light propagates with the wave vector $k$. The scattered fields propagate with the wave vector $k_z = k/cos(\alpha_0)$, which depends on the nanofiber diameter. Here $\alpha_0$ is the polar angle of the incident evanescent field of the fiber. The total forces acting on the particles experience a long-range modulation with a period close to $\lambda_l = 2\pi/(k-k_z)$ when the scattered light propagates along the fiber, in the same direction as the guided light. A short-range modulation with a period close to $\lambda_s = 2\pi/(k+k_z)$ results from the interference of the evanescent wave and the scattered light propagating in opposite direction to the guided light [21].

For the numerical optical force calculation, we first simulate the forces acting on a single particle, adjacent to a nanofiber, in water. Figure 1 shows the scattering force, $F_s$ (blue), and gradient force, $F_g$ (red), on a 3.13 μm $SiO_2$ sphere as a function of nanofiber diameter. The scattering and gradient forces have maximum values for fiber diameters of 470 nm and 600 nm, respectively. Both forces become weaker when the fiber diameter increases beyond these values. The decrease of the forces is due to the $E$-field at the interface which decreases with increasing waist diameter. Furthermore, when the tapered fiber diameter is below a certain value a considerable fraction of the guided power leaks out from the fiber interface into the surrounding medium. This, in turn, leads to a sharp decrease of the scattering and gradient forces.

To investigate optical binding effects, the forces on two neighboring 3.13 μm $SiO_2$ particles were calculated. The nanofiber diameter was chosen to be 550 nm, since for this parameter the scattering force is at a maximum while maintaining a strong trapping gradient force. Figure 2(a) shows two particles (indicated by $P_1$ and $P_2$) in the nanofiber evanescent field as the incident light field propagates from the left-hand side. Each sphere ($P_1$ or $P_2$) experiences a trapping gradient force directed towards the nanofiber surface ($F_{g\text{-}1}$ or $F_{g\text{-}2}$, respectively), and a scattering force in the $x$-direction of light propagation ($F_{s\text{-}1}$ or $F_{s\text{-}2}$, respectively). For a two-sphere system, the induced scattering and modification of the evanescent field in the wake of each sphere is evident. It is also clear that the optical forces on $P_1$ and $P_2$ are influenced by both incident evanescent and scattered fields from the particles, and a proportion of the scattered light is recoupled into the fiber. By modifying the relative distance between the spheres, the magnitudes of the scattering and gradient forces on $P_1$ and $P_2$ are calculated for interparticle separations up to 45 μm, as illustrated in Fig. 2(b).

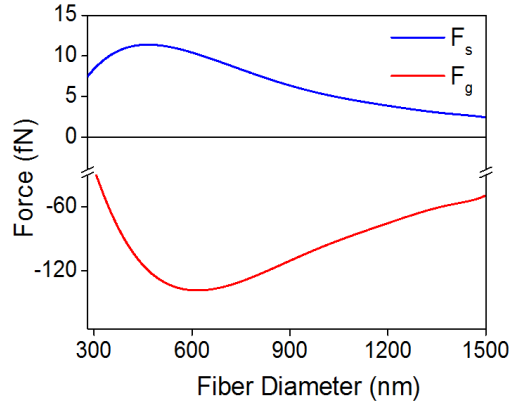

Fig. 1. Gradient and scattering forces acting on a single 3.13 μm particle as a function of fiber diameter.

The large amplitude and low frequency oscillations in the scattering and gradient forces on $P_2$ are due to the long-range binding effect instigated by the interference of the nanofiber evanescent wave with the forward-scattered field from particle $P_1$. High frequency oscillations are observed in the forces on $P_1$ and these are a signature of the short range binding effects, which are caused by the interference of the incident wave and back-scattered field from the surface of $P_2$. It is worth noticing that these short range binding oscillations are relatively independent of the interparticle distance, though this may seem counterintuitive. We attribute this to the fact that the backscattered field from $P_2$ is coupled back to the fiber, and thus can be transferred, without damping, over arbitrary distances. However, similar simulations for free space optical binding show clear attenuation of the oscillations with particle separation (see Fig. 1 in [22]).

The scattering force on both particles is positive (in the direction of incident light), thus the spheres propel along the fiber. The negative gradient force represents the attractive trapping force towards the fiber surface.

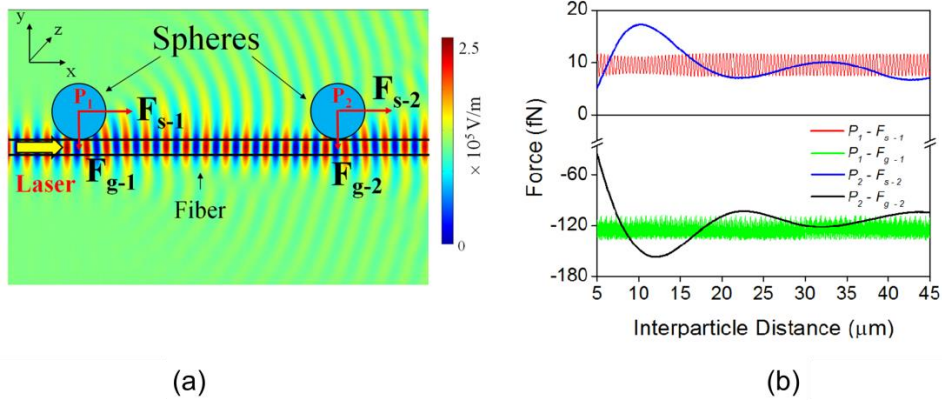

Fig. 2. (a) Gradient and scattering ($F_g$, $F_s$) forces on nearby particles in the nanofiber evanescent field. (b) Magnitude of gradient and scattering forces on two 3.13 μm particles in the evanescent field of a 550 nm nanofiber.

The optical binding force between the spheres is found by calculating the difference between the scattering forces, i.e. $F_{s-2} - F_{s-1}$. This binding force is depicted in Fig. 3(a), along with its associated calculated potential. Points of equilibrium in the system occur when the total optical binding force is zero and has a negative slope. To illustrate this in detail, Fig. 3(b) is focused on the first zero crossing of Fig. 3(a). Combining this information with that of the overall potential, the favored particle positions for system equilibrium may be identified. In this particular sphere-fiber combination, there are two equilibrium binding distances – a much stronger primary equilibrium at 17 μm interparticle spacing and a secondary one at 36 μm.

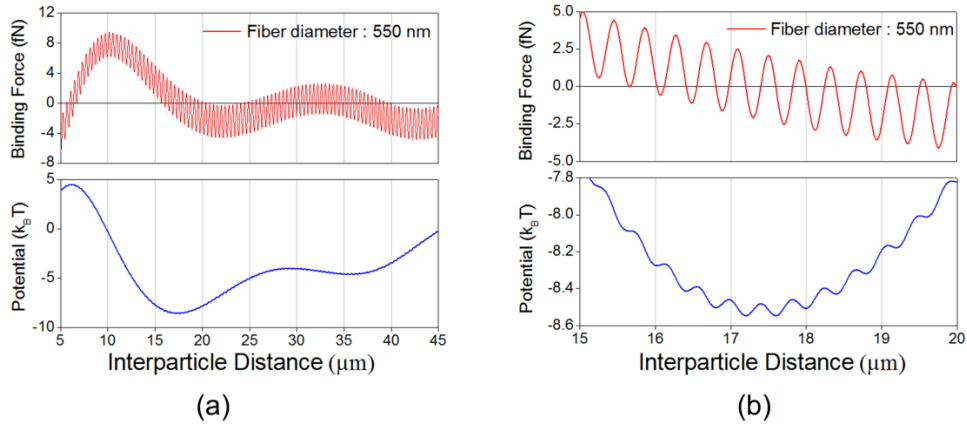

Fig. 3. (a) Calculated binding force (top) and potential (bottom) for two 3.13 μm particles adjacent to a 550 nm diameter nanofiber, (b) Zoom of plot (a) for inter particle distances of 15-20 μm.

3.2 *Effect of particle size and fiber diameter on optical binding force*

Similar models are used to analyze the effects of varying particle size and nanofiber diameter on the interparticle binding distances. In Fig. 4(a), the fiber diameter was fixed at 550 nm - the solid line corresponds to the binding force between two 2.03 μm particles and the dotted line corresponds to a pair of 3.13 μm particles. The smaller particles clearly have a shorter force modulation period. The diameter of the nanofiber also plays a strong role in the nature of this interaction [Fig. 4(b)]. In fact, larger particles or decreasing nanofiber diameters both cause a relative downward shift of the binding force graph.

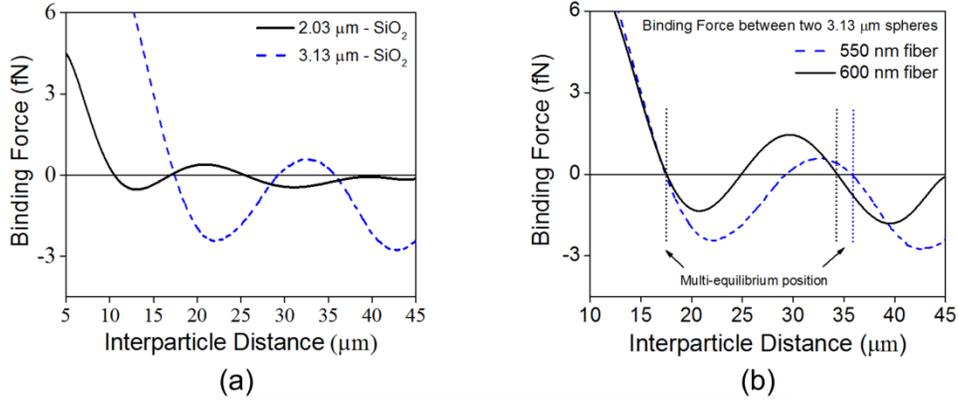

Fig. 4. (a) Binding force between two 3.13 μm (dashed) and two 2.03 μm (solid) particles. (b) Binding forces between two 3.13 μm spheres in the evanescent field of a nanofiber of diameter 550 nm (dashed) and 600 nm (solid).

The dependence of equilibria positions and the shape of associated potentials on the nanofiber diameter are further illustrated in Fig. 5(a). There is clear evidence of a lateral shift of the first and, in particular, the second equilibria distances as the nanofiber diameter reduces. Significantly, the depth of the potential of the binding interaction for the second equilibrium point reduces with decreasing fiber diameter, coupled with a marked increase in the depth at the first point. Thus, for smaller fibers, the second particle of a pair will more likely be bound at the first equilibrium point. In fact, for nanofibers that are smaller again, the curve only crosses the equilibrium line once, resulting in a single preferable binding length.

Figure 5(b) compares simulated interparticle distance over a range of fiber diameters for both 2.03 and 3.13 μm spheres. This graph has some noteworthy features. Firstly, the dotted lines signify the result modelled by assuming uniform illumination of the evanescent field on both spheres; the solid lines recognize the loss of power following scattering from $P_1$. Secondly, the percentage of transmitted light present in the evanescent field generally increases as the size of the nanofiber decreases [6]. Thus for larger fibers there is little scattering by $P_1$. It follows that the deviation from the dotted curve only arises for smaller fibers whose intense evanescent field undergoes more scattering. This deviation is more pronounced, and occurs at a larger fiber diameter, for the 3.13 μm sphere – the particle's increased size gives it a greater presence in the field, leading to enhanced scattering. The marked points on the plot correspond to sample experimental data recorded in the manner described in Section 4.

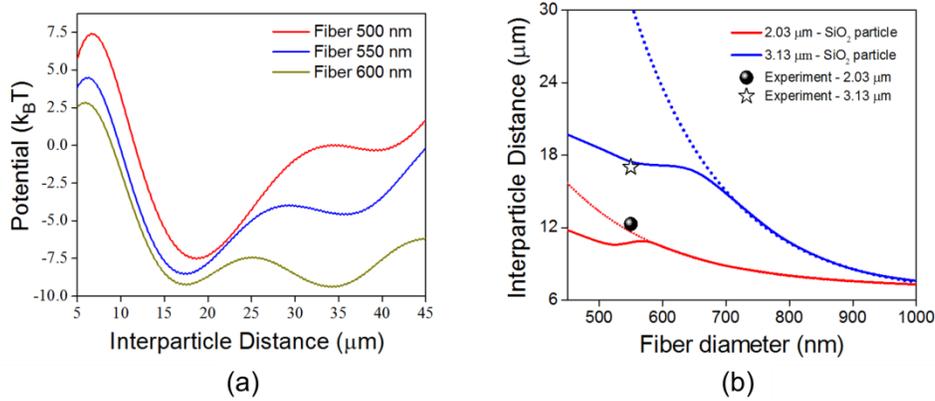

Fig. 5. (a) Optical binding potential between two 3.13 μm spheres as a function of nanofiber diameter. (b) Plot showing the influence of nanofiber diameter on the position of the first equilibrium point between two spheres (red 2.03 μm and blue 3.13 μm). The dotted plots assume uniform evanescent illumination of both spheres; solid lines account for scattering by the first particle.

## 4. Experiment

*4.1 Optical nanofiber fabrication and integration into optical tweezers*

Optical nanofibers may be produced in the laboratory by a variety of techniques [18, 19]. Using a custom-built, hydrogen flame-brushed rig [20], we routinely fabricate linearly-tapered nanofibers with diameters in the range 450 - 1000 nm and transmissions of up to 99% for the fiber fundamental mode. The nanofibers were integrated into a home-built optical tweezers based on Thorlabs Kit (OTKB/M, λ = 1064 nm) with a 100 X 1.25 NA oil immersion objective, as is commonly used to trap particles in liquid dispersion through a microscope cover slip. The working distance allowed repositioning of the trapped particles up to 100 μm above the slide surface. To facilitate accurate positioning of the nanofiber in the proximity of this trap, it was affixed to a custom-built mount post fabrication; this provided vertical positioning and horizontal tilt adjustment. Figure 6 (a) depicts a schematic illustrating this integration scheme. Further details of the setup are given in [15].

The combined 'nanofiber and tweezers' system was used to study the optical binding effects on 3.13 μm $SiO_2$ particles in the nanofiber evanescent field. Figure 6(b) and 6(c) show a 3.13 μm sphere trapped via the tweezers close to, and then at, the nanofiber surface, respectively. This particle transfer may be done by moving the sample chamber via translation or by shifting the position of the trap through the objective by galvo-mirror pair adjustment. If light propagated through the nanofiber, the particle introduction was detected via a pronounced and repeatable fiber transmission dip. The optical power in the nanofiber was stable to within ± 0.5%. The trap can be time-shared via galvo-steered mirrors to harness a number of particles simultaneously and introduce them to the fiber. This method was exploited to investigate binding effects and speed variations of chains of one to seven particles (Sec. 4.2.1-2 below).

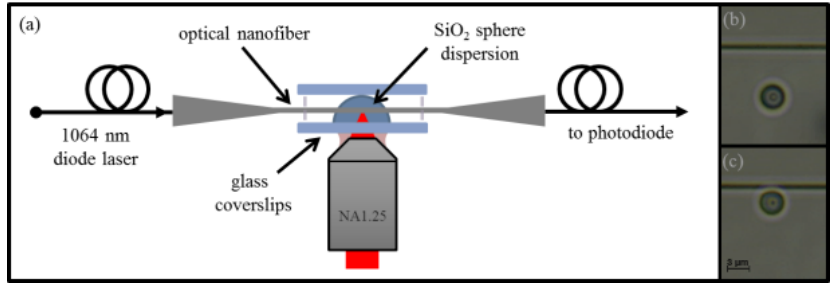

Fig. 6. (a) Schematic of optical nanofiber mounted within an optical tweezers. (b) 3.13 µm sphere trapped close to the nanofiber surface. (c) Trapped sphere at the surface of the nanofiber.

*4.2 Experimental results*

*4.2.1 Particle chains in the nanofiber evanescent field*

To experimentally investigate the evanescent field optical binding as outlined above, the integrated optical 'nanofiber and tweezers' system was used to introduce particles to the nanofiber in chains of increasing length. The trapping mechanism was then switched from the standard optical tweezers to nanofiber-mediated trapping and the ensuing interaction observed and recorded. Care was taken to ensure that all spheres were removed from contact with the fiber between runs to avoid unwanted scattering field affects. We repeated this systematic introduction and propulsion exercise for particle chains of length one to seven particles (see Fig. 7, top to bottom). The observed particle dynamics displayed clear trends of particle positioning. To analyze these chains, the particles were labelled $P_a$ to $P_g$ and the interparticle distances $d$ were calculated for each pairing. Figure 7 is a sequence of images showing the increasing number of trapped particles at the surface of the nanofiber.

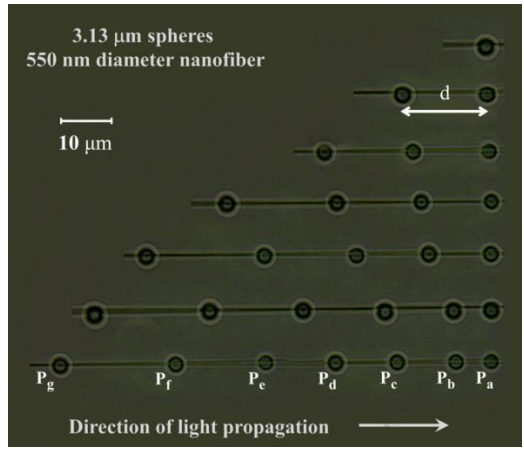

Fig. 7. Typical experimental observation of particle chains of increasing length from one to seven particles, self-arranged under propulsion. Interparticle distance is labelled $d$ and individual spheres are labelled from right to left, $P_{a-g}$. (Note: Disjoint in fiber focus for composite images of 4 to 7 particles due to restricted field of view of objective).

In Fig. 8(a), the chains are plotted as a function of interparticle distance, $d$, with different colors representing each chain. This clearly shows that the distance between neighboring particles in every chain increases consistently, from the front particle ($P_a$) to the back particle ($P_g$), where the back particle is the most upstream particle in each chain with respect to the propagation direction. Similar effects have been observed in various systems of standing wave optical binding [21], although in these cases the spacing effects were symmetric around the central particle due to the nature of the counter-propagating fields. Note that for the two-sphere chain, the measured value for $d$ (~17 μm) is close to that predicted in Fig. 5(b) for 3.13 μm $SiO_2$ spheres and a 550 nm diameter fiber.

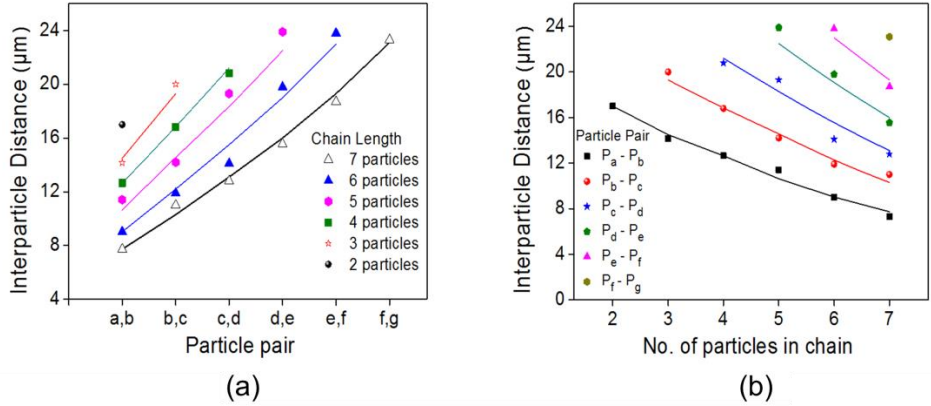

Fig. 8. (a) Plot of $d$ in increasing chain lengths as a function of specific pairings. This indicates increasing $d$ values from the front to the back of a given chain. (b) Inter-particle distance plotted against number of particles in a chain for a given pair progression. Here, it is clear that, in every case, the distance between a given pair decreases as the chain length increases. Color lines in both (a) and (b) are trend lines to guide the eye.

Figure 8(b) shows a second trend evident in the data; as the number of spheres in a chain increases, the distance between comparable sphere pairs in subsequent chains decreases. To give a specific example, as the number of particles in a chain increases, the inter-particle distance between $P_a$ and $P_b$ decreases.

Both trends can be understood qualitatively if we consider an auxiliary simulation based on the model shown in Fig. 2. Instead of keeping the particle sizes identical, we calculated the distance between two particles of different sizes. In this model, the diameter of the second sphere ($P_2$ on Fig. 2(a)) was fixed, and the diameter of $P_1$ was gradually increased. The results are shown in Fig. 9, and indicate that the interparticle distance decreases as the first sphere becomes bigger. As this bigger sphere $P_1$ will scatter more light than a smaller sphere of similar material, it is intended to represent the increasing scattering from the growing number of upstream particles. This higher scattering causes the scattering force curve of particle $P_1$ (red curve in Fig. 2(b)) to shift upwards, thus effectively shifting the binding curve downwards. Finally, the intersection of the binding force curve with the zero-equilibrium line [Fig. 3(a)] occurs at a shorter $d$. This would imply two consequences: firstly, in longer chains, the compound scattering effects of two or multiple spheres on the rest of the chain causes a reduction in subsequent $d$ values [Fig. 8(b)]; and secondly, within a single chain, the more upstream spheres there are before a given sphere, the shorter the distance between this sphere and the upstream particles [Fig. 8(a)].

Finally, it is interesting to note that, occasionally during propulsion, a slight defect in the fiber caused the particles to migrate around the fiber. In this event, all particles moved over in quick succession, and continued to propel while maintaining their previously spaced array. Also, non-uniformities in nanofiber diameter or surface flaws can sometimes cause momentary oscillations in the binding distances within a longer chain. In such cases, there is a spring-like effect - if one particle (e.g. the third) is held back slightly, or accelerated, due to directional scattering from a surface flaw, its neighbor will compensate in position and vice versa, until the chain reaches equilibrium (e.g. Fig. 8, blue dataset for six particles). These observations serve as further evidence for optical binding effects between the spheres.

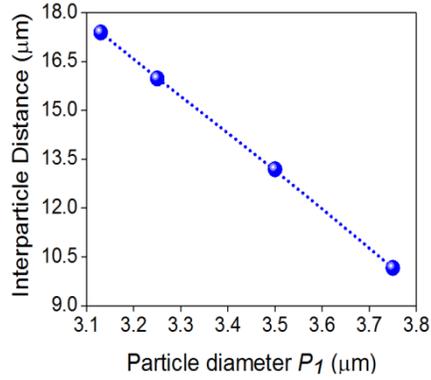

Fig. 9. Simulated data of reducing equilibrium binding distance between $P_1$ and $P_2$, as the size of $P_1$ is gradually increased from 3.13 μm. The solid line is a trend line to guide the eye.

*4.2.2 Speed variations with number of particles in bound chain*

Repeated experiments of the nature outlined above showed enhanced particle speeds in chains when compared with those of single particles under the same illumination power. Figure 10 shows a sequence of video frames taken at 0.2 s intervals of three 3.13 μm bound particles being propelled along the tapered region of the fiber.

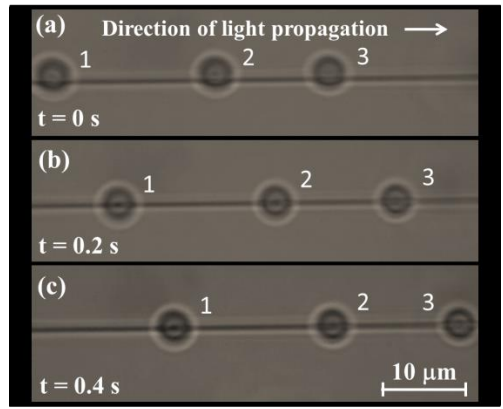

Fig. 10. Sequence of video frames showing three 3.13 μm bound particles being propelled along the tapered region of the fiber. One can see that between the (a) and (b) frames, all the particles simultaneously shifted downwards with respect to the fiber, along with propelling to the right.

The speed of the 3.13 μm particles was recorded in groupings of one to six, under 30 mW input light in the nanofiber. Here, the grouped particles propelled at speeds averaging 25% faster than single spheres. Similar increased speeds of two-sphere groupings were reported for particles under evanescent-field counter-propagation with ridge waveguides [23].

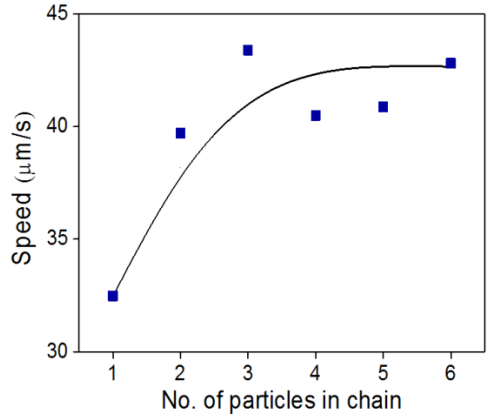

Fig. 11. Speeds of 3.13 μm particle chains of increasing length, with 30 mW input power in a nanofiber of 550 nm diameter, with included data trend-line.

The observed results in Fig. 11 indicate that the optical binding effect also leads to the enhancement of bound particle velocities. Each particle within the chain is optically coupled to each of the other particles. This results in the enhancement of coupling evanescent light from the nanofiber to the particle chains. This motional enhancement with multi-particle systems has implications for future design and calibration of nanofiber-based particle delivery schemes.

## 5. Discussion

The presented results demonstrate use of a simple, two-particle simulated model to investigate the stable configurations of one-dimensional longitudinal particle interactions in the evanescent field of a nanofiber. The influence of particle size and nanofiber diameter on inter-particle separation shows good agreement with experimental observations. We found that by reducing the fiber diameter, the evanescent field at the nanofiber surface increased which in turn increased the scattering light acting upon each individual particle. This results in the enhancement of the optical binding effect for self-assembled chains of particles along a nanofiber. A similar phenomenon was also observed for larger particles. Our theoretical and experimental results indicate that stable configurations for trapping micro-objects in the evanescent field of nanofibers with diameters less than 550 nm tend to be bound at the closer equilibrium positions.

In the case of a long chain, multi-particle trap, the optical power changes along the nanofiber due to the propagation and scattering loses. This leads to the modification of bound structures of particles within the chains. The physical meaning of the obtained effect is assigned changes in the interference of the incident and scattered fields from particles along the nanofiber. In the presented work, the experiments conducted included up to seven particles. However, when the particle number was further increased, the particle speed reduced as the last few particles (see Fig. 7) within the chains stuck together leading to an unstable bound structure. The aim of this paper is to focus on the far-field binding effect of micro-objects, so such a case of unstable bound structures is outside its scope.

Furthermore, as shown in Fig. 11, experimental results suggest that the particle speed increases with the number of particles (up to six particles) in the bound structure. To obtain a better understanding of this phenomenon, the hydrodynamic interaction between the surrounding fluid and particles should be considered. Multiple particles moving in a fluid excite flows through the no-slip boundary condition at their surfaces. These flows result in an increase of particle velocity since the surrounding fluid displaced by one particle favors the movement of others. Additionally, the presence of the fiber needs to be accounted for, as it retards the motion of nearby particles. It is worth noting that relative stable configurations of a particle chain may also be influenced by the enhancement of the hydrodynamic coupling between particles. To reduce these effects, the experiments were conducted using low laser powers (10 – 50 mW), so the interparticle distance did not change for a fixed number of particles.

A quantitative explanation of our experimental observation in Fig. 10 requires a solution to Maxwell's equations for a chain structure of more than two particles close to the surface of the nanofiber. This will provide further information on the effect of multiple scattering on the involved particles.

**6. Conclusion**

Optical binding of micro-particles in a nanofiber evanescent field was theoretically and experimentally studied in this paper. Such knowledge of significant interparticle optical interactions and associated chain-related speed effects is essential for future studies on trapping geometries in the nanofiber evanescent field, including general applications of nanofibers for long-range and large scale particle conveyance [10, 11], counter propagation and controlled standing wave trapping [24], and particle dynamics under higher mode propagation [25]. Above all, the presented results demonstrate the versatility of the integrated optical nanofiber/optical tweezers system for selective particle optical binding investigations, with simply modulated, highly tunable parameters including transmitted wavelength and nanofiber diameter.

**Acknowledgments**

This work is supported by OIST Graduate University and Science Foundation Ireland under Grant No. 08/ERA/I1761 through the NanoSci- E+ Transnational Programme. MCF and SNC acknowledge support through COST Action MP0604. MS acknowledges the support of JSPS through the Postdoctoral Fellowship for Overseas Researchers scheme.